\documentstyle[12pt,epsfig]{article}

\newcommand{\beq}{\begin{equation}}
\newcommand{\eeq}{\end{equation}}
\newcommand{\barr}{\begin{eqnarray}}
\newcommand{\earr}{\end{eqnarray}}

\newcommand{\andy}[1]{ }

\def\cC{{\cal C}}
\def\cZ{{\cal Z}}
\def\As{{\cal A}}

\begin{document}

\begin{titlepage}
\begin{flushright}
\today \\
BA-TH/99-336\\
\end{flushright}
\vspace{.5cm}
\begin{center}
{\LARGE Deviations from exponential law and Van Hove's
``$\lambda^2t$" limit }

\quad

{\large P. FACCHI\footnote{email: Paolo.Facchi@ba.infn.it}
        and S. PASCAZIO\footnote{email: Saverio.Pascazio@ba.infn.it}\\
           \quad    \\

        Dipartimento di Fisica, Universit\`a di Bari

        and Istituto Nazionale di Fisica Nucleare, Sezione di Bari \\
 I-70126  Bari, Italy \\

}

\vspace*{.5cm} PACS numbers: 05.40.-a; 31.70.Hq, 03.65.-w; 31.30.Jv
\vspace*{.5cm}

{\small\bf Abstract}\\ \end{center}

{\small The deviations from a purely exponential behavior in a
decay process are analyzed in relation to Van Hove's ``$\lambda^2
t$" limiting procedure. Our attention is focused on the effects
that arise when the coupling constant is small but nonvanishing. We
first consider a simple model (two-level atom in interaction with
the electromagnetic field), then gradually extend our analysis to a
more general framework. We estimate all deviations from exponential
behavior at leading orders in the coupling constant.}

\vspace*{1cm} Keywords: Exponential Law, Fermi Golden Rule, Markovianity

\end{titlepage}

\newpage

\setcounter{equation}{0}
\section{Introduction  }
\label{sec-introd}
\andy{intro}

The evolution law in quantum mechanics is governed by unitary operators
\cite{Dirac}. This entails, by virtue of very general
mathematical properties, that the decay of an unstable quantum system cannot
be purely exponential.
In general, a rigorous analysis based on the Schr\"odinger equation shows that
the decay law is quadratic for very short times \cite{shortt} and
governed by a power law for very long times \cite{longtt}.
These features of the quantum evolution are well known and
discussed in textbooks of quantum mechanics \cite{Sakurai} and
quantum field theory \cite{Brown}.
The temporal behavior of quantum systems is reviewed in Ref.\ \cite{temprevi}.

Although the domain of validity of the exponential law is limited,
the Fermi ``Golden Rule" \cite{Fermigold} works very well and no
deviations from the exponential behavior have ever been observed
for truly unstable systems \cite{expttlaw}. The quantum mechanical
derivation of this law is based on the sensible idea that the
temporal evolution of a quantum system is dominated by a pole near
the real axis of the complex energy plane (Weisskopf-Wigner
approximation \cite{seminal}). This yields an irreversible
evolution, characterized by a master equation and exponential decay
\cite{VanKampen}. An important contribution to this issue was given
in the 50's by Van Hove \cite{vanHove}, who rigorously showed that
it is possible to obtain a master equation (leading to exponential
behavior) for a quantum mechanical system endowed with many
(infinite) degrees of freedom, by making use of the so-called
``$\lambda^2 t$" limit. The crucial idea is to consider the limit
\andy{vHlimit}
\beq
\lambda\to0 \quad\mbox{keeping}\quad \tilde t =
\lambda^2t\quad\mbox{finite ($\lambda$-independent constant)},
\label{eq:vHlimit}
\eeq
where $\lambda$ is the coupling constant and $t$ time. One then
looks at the evolution of the quantum system as a function of the
rescaled time $\tilde t$. There has recently been a renewed
interest in the physical literature for this time-scale
transformation and its subtle mathematical features: see
\cite{Accardi}.

The purpose of this paper is to consider the effects that arise
when the coupling constant is small but nonvanishing. This will
enable us to give general estimates for deviations from exponential
behavior. The  paper is organized as follows. We shall first look,
in Section 2, at a simple system: we summarize some recent results
on a characteristic transition of  the hydrogen atom in the
two-level approximation. In Section 3 we consider the action of the
Van Hove limiting procedure on a generic two-level atom in the
rotating-wave approximation and then generalize our result when the
other discrete levels and the counter-rotating terms are taken into
account. We look in particular at the scaling procedure from the
perspective of the complex energy plane, rather than in terms of
the time variable. This enables us to pin down the different
sources of non-exponential behavior. In Section 4 our analysis is
extended to a general field-theoretical framework: general
estimates are given of all deviations from the exponential law
(both at short and long times) at leading orders in the coupling
constant.

\setcounter{equation}{0}
\section{Hydrogen atom in the two-level approximation }
\label{sec-2levels}
\andy{2levels}

We start our considerations from a simple field-theoretical model.
Consider the Hamiltonian ($\hbar=c=1$)
\andy{tothaml1,2,3}
\barr
 H & = & H_0 + \lambda V ,
       \label{eq:tothaml1} \\
 H_0 & = & H_{\rm atom} + H_{\rm EM} \nonumber \\
     & \equiv & \omega_0 b^\dagger_2 b_2 + \sum_\beta
\int_0^\infty d\omega \,
\omega a^\dagger_{\omega\beta} a_{\omega\beta} , \label{eq:tothaml2} \\
 V & = & \sum_\beta \int_0^\infty
d\omega \left[ \varphi_\beta(\omega) b^\dagger_1 b_2 a^\dagger_{\omega \beta}
+ \varphi^*_\beta(\omega) b^\dagger_2 b_1 a_{\omega \beta}
\right], \label{eq:tothaml3}
 \earr
where $H_{\rm atom}$ is the free Hamiltonian of a two-level atom
($\omega_0$ being the energy gap between the two atomic levels),
$b_j, b_j^\dagger$ are the annihilation and creation operators
of the atomic level $j$, obeying anticommutation relations
\andy{fermicomm}
 \beq
 \{b_k, b^\dagger_\ell \} = \delta_{k\ell}  \qquad (k,\ell=1,2),
 \label{eq:fermicomm}
 \eeq
$H_{\rm EM}$ is the Hamiltonian of the free EM field, $\lambda$ the coupling
constant and $V$ the interaction Hamiltonian.
We are working in the rotating-wave approximation
and with the energy-angular momentum basis for photons \cite{Heitler}, with
$\sum_\beta = \sum_{j = 1}^\infty\sum_{m=-j}^j\sum_{\epsilon = 0}^1
$, where $\epsilon$
defines the photon parity $P=(-1)^{j+1+\epsilon}$, $j$ is the total
angular momentum (orbital+spin) of the photon, $m$ its magnetic
quantum number and
 \andy{boscomm2}
 \beq
 [a_{\omega j m \epsilon}, a^\dagger_{\omega' j' m'\epsilon'}] =
\delta(\omega-\omega')\delta_{j j'}\delta_{m m'}\delta_{\epsilon\epsilon'}.
 \label{eq:boscomm2}
 \eeq
We shall focus our attention on the 2P-1S transition of hydrogen, so that
$\omega_0 = \frac{3}{8} \alpha^2 m_e \simeq 1.550 \cdot 10^{16}$ rad/s
($\alpha$ is the fine structure constant and $m_e$ the electron
mass) and the matrix elements $\varphi_{\beta}(\omega)$ of the interaction
are known exactly \cite{Moses,Seke}
 \andy{phidef1}
 \barr
\varphi_\beta (\omega) &=& \langle 1; 1_{\omega\beta}|V|2;0\rangle =
\varphi_{\bar{\beta}}(\omega)\delta_{\beta\bar{\beta}}\nonumber\\
&=& i(\Lambda)^\frac{1}{2}\frac{\left(
\frac{\omega}{\Lambda}\right)^\frac{1}{2}}{\left[1 + \left(
\frac{\omega}{\Lambda}\right)^2 \right]^2}
\delta_{j1}\delta_{mm_2}\delta_{\epsilon 1},
 \label{eq:phidef1}
 \earr
where
\andy{stati}
 \beq
 |1; 1_{\omega\beta}\rangle  \equiv |1\rangle \otimes |\omega, j, m,
\epsilon \rangle, \quad
 |2; 0\rangle \equiv |2\rangle \otimes |0\rangle   \label{eq:stati}
 \eeq
(the first ket refers to the atom and the second to the photon) and
the selection rule, due to angular momentum and parity
conservation, entails that the only nonvanishing term in
(\ref{eq:tothaml3}) and (\ref{eq:phidef1}) is $\bar{\beta} = (1,
m_2, 1)$. We emphasize that the so-called "retardation effects" are
taken into account in (\ref{eq:phidef1}). The normalization reads
\andy{normaliz}
 \beq
\langle 1; 1_{\omega\beta} |1; 1_{\omega'\beta'}\rangle =
\delta (\omega-\omega') \delta_{\beta\beta'}, \quad
\langle 2;0 |2; 0\rangle =1
\label{eq:normaliz}
 \eeq
and the quantities
 \andy{lambdachi}
 \barr
 \Lambda &=& \frac{3}{2} \alpha m_e = \frac{3}{2a_0}
\simeq 8.498\cdot 10^{18} \mbox{rad/s},
 \nonumber \\
 \lambda &=& \left( \frac{2}{\pi} \right)^{1/2}
\left(\frac{2}{3}\right)^{9/2} \alpha^{3/2} \simeq .802 \cdot 10^{-4},
 \label{eq:lambdachi}
 \earr
are the natural cutoff of the atomic form factor,
expressed in terms of the Bohr radius $a_0$, and the coupling
constant, respectively. Observe that there are no free parameters in
(\ref{eq:tothaml1})-(\ref{eq:tothaml3}).

The above model was analyzed in a previous paper \cite{FP1}, where
we mainly concentrated our attention on the deviations from
exponential, both at short and long times. There is interesting
related work on this subject \cite{KnightMilonni'76}. Let us
summarize the main results, by concentrating our attention on the
role played by the coupling constant $\lambda$.

Assume one can prepare (at time $t=0$) the atom in the initial
state $|2;0\rangle$. This is an eigenstate of the unperturbed
Hamiltonian $H_0$, whose eigenvalue is $\omega_0$. The evolution is
governed by the unitary operator $U(t) = \exp (-iHt)$ and the
``survival" or nondecay amplitude and probability at time $t$ are
defined as (interaction picture)
 \andy{survampl,ndq1}
 \barr
 \As(t) & = & \langle 2;0|e^{i H_0 t}U(t)|2;0\rangle ,
\label{eq:survampl} \\
P(t) & = & |\langle 2;0 |e^{i H_0 t}U(t) |2;0 \rangle |^2 .
\label{eq:ndq1}
\earr
The survival probability at short times reads
\andy{ndq, naiedef}
\beq
P(t) = 1 - t^2/\tau_{\rm Z}^2 + \cdots,
\qquad \tau_{\rm Z} \equiv (\lambda^2\langle 2;0|V^2|2;0\rangle)^{-1/2} .
\label{eq:naiedef}
\eeq
The quantity $\tau_{\rm Z}$ is the so-called ``Zeno time"
and yields a quantitative estimate of the deviation from exponential at very
short times. Strictly speaking, $\tau_{\rm Z}$ is the convexity
of $P(t)$ in the origin. One finds \cite{FP1}
\andy{taudet}
\beq
 \tau_{\rm Z} = \frac{\sqrt{6}}{\lambda\Lambda} =
(3 \pi)^{\frac{1}{2}}
\left(\frac{3}{2}\right)^{\frac{7}{2}} \frac{1}{\alpha^{\frac{5}{2}} m_e}
\simeq 3.593 \cdot 10^{-15} \mbox{s}.
 \label{eq:taudet}
 \eeq
It is possible to obtain a closed expression for $\As(t)$, valid at
all times, as an inverse Laplace transform:
\andy{survamp1,qs3}
 \barr
 \As(t) &=& \frac{e^{i\omega_0 t}}{2 \pi i}
\int_{\rm B} ds \frac{e^{s \Lambda t}}{s + i
\frac{\omega_0}{\Lambda} + \lambda^2 Q(s)},
 \label{eq:survamp1} \\
 & & \quad Q(s) \equiv -i \int_0^\infty dx \frac{x}{(1+x^2)^4} \frac{1}{x-is} ,
 \label{eq:qs3}
 \earr
where B is the Bromwich path, i.e.\ a vertical line at the right of
all the singularities of the integrand. $Q$ is a self-energy
contribution and can be computed exactly:
 \andy{qs4}
 \barr
 Q(s) &=& \frac{-15\pi i -(88-48\pi i)s - 45\pi is^2 + 144s^3}
 {96(s^2 - 1)^4}\nonumber\\
 & &+\frac{15\pi is^4- 72s^5 - 3\pi is^6 + 16s^7-96 s\log s}{96(s^2 - 1)^4} .
 \label{eq:qs4}
 \earr
At short and long times one gets
 \andy{shortt,largetimes1}
 \barr
\!\! P(t) \! &\sim& \!
1 - \frac{t^2}{\tau_{\rm Z}^2} \qquad (t\ll \tau_{\rm Z}),
 \label{eq:shortt} \\
\!\! P(t) \! &\sim& \!
\cZ^2 e^{-\gamma t}+ \lambda^4\frac{\cC^2}{(\omega_0 t)^4}
- 2 \lambda^2\frac{\cC\cZ}{(\omega_0 t)^2} e^{-\frac{\gamma}{2} t}
 \cos\left[(\omega_0-\Delta E) t -\zeta \right] \label{eq:largetimes1} \\
  & & \qquad\qquad\qquad\qquad\qquad\qquad\qquad\qquad(t\gg \Lambda^{-1}),
  \nonumber
 \earr
where
 \andy{fgr, shift,Zz,Cchi}
 \barr
 \gamma &=& 2\pi \lambda^2|\varphi_{\bar{\beta}}(\omega_0)|^2 +
{\rm O}(\lambda^4)
= 2\pi\lambda^2\omega_0 +{\rm O}(\lambda^4)
\simeq 6.268\cdot10^8\mbox{s}^{-1},
 \label{eq:fgr}\nonumber\\ \\
\Delta E &=&\lambda^2 {\cal P}\!
\!\int_0^\infty d\omega |\varphi_{\bar{\beta}}(\omega)|^2
\frac{1}{\omega -\omega_0}+{\rm O}(\lambda^4)\simeq 0.5 \lambda^2 \Lambda,
 \label{eq:shift} \\
\cZ e^{i\zeta} &\simeq& (1-4.38\lambda^2)e^{-i1.00\pi\lambda^2}=1 + {\rm O}(\lambda^2),
 \label{eq:Zz} \\
\cC & \simeq& 1+5.38\lambda^2 = 1 + {\rm O}(\lambda^2).
 \label{eq:Cchi}
 \earr
The first two formulae give the Fermi ``Golden Rule," yielding the lifetime
\andy{fgrule}
 \beq
\tau_{\rm E}=\gamma^{-1} \simeq 1.595\cdot 10^{-9}\mbox{s},
 \label{eq:fgrule}
\eeq
and the second order correction to the energy level $\omega_0$.
The exact expressions for the quantities
(\ref{eq:fgr})-(\ref{eq:Cchi}) are given in Ref.\ \cite{FP1}.

\setcounter{equation}{0}
\section{Van Hove's limit }
\label{sec-VHL}
\andy{VHL}

Let us look at Van Hove's $``\lambda^2 t"$ limiting procedure
applied to the model of the previous section. Before proceeding to
a detailed analysis, it is worth putting forward a few preliminary
remarks: we  shall scrutinize (in terms of the coupling constant)
the mechanisms that make the nonexponential contributions in
(\ref{eq:shortt})-(\ref{eq:largetimes1}) vanish. To this end,
observe first that as $\lambda \to 0$ the Zeno time
(\ref{eq:taudet}) diverges, while the {\em rescaled} Zeno time
vanishes
\andy{rescz}
\beq
\tilde\tau_{\rm Z} \equiv \lambda^2\tau_{\rm Z}=\lambda\frac{\sqrt{6}}{\Lambda}
={\rm O}(\lambda).
\label{eq:rescz}
\eeq
On the other hand, the rescaled lifetime (\ref{eq:fgrule}) remains
constant [see (\ref{eq:fgr})]:
\andy{resce}
\beq
\tilde\tau_{\rm E} \equiv \lambda^2\tau_{\rm E}=\frac{1}{2\pi\omega_0}={\rm O}(1).
\label{eq:resce}
\eeq
Moreover, the transition to a power law occurs when the first two
terms in the right hand side of (\ref{eq:largetimes1}) are
comparable, so that
\andy{nuovapot}
\beq
(\omega_0 t)^2e^{-\frac{\gamma}{2}t}\simeq\lambda^2,
\label{eq:nuovapot}
\eeq
because both $\cC$ and $\cZ$ are $\simeq 1$. In the limit of small
$\lambda$, (\ref{eq:nuovapot}) yields $t = \tau_{\rm pow}$, with
\andy{interr1}
\beq
2 \log (\omega_0 \tau_{\rm pow}) - \frac{\gamma}{2}\tau_{\rm pow}
\simeq 2 \log \lambda,
\label{eq:interr1}
\eeq
namely, by (\ref{eq:fgr}),
\andy{interr2}
\beq
\frac{\tau_{\rm pow}}{\tau_{\rm E}} \simeq
 4\log \frac{1}{\lambda} + 4\log \frac{\tau_{\rm pow}}{2\pi \lambda^2
\tau_{\rm E}} =
 12\log \frac{1}{\lambda} + 4\log \frac{\tau_{\rm pow}}{\tau_{\rm E}}
 + 4 \log \frac{1}{2\pi}.
\label{eq:interr2}
\eeq
Therefore, when time is rescaled,
\andy{rescpl}
\beq
\tilde\tau_{\rm pow} \equiv \lambda^2\tau_{\rm pow}=12\tilde\tau_{\rm E}
\log\frac{1}{\lambda}+{\rm O}\left(\log\log\frac{1}{\lambda} \right)=
{\rm O}\left(\log\frac{1}{\lambda}\right) .
\label{eq:rescpl}
\eeq
Finally, the power contributions are $\sim {\rm
O}(\lambda^{3\alpha})\tilde t^{-\alpha} \,(\alpha=2,4)$, the period
of the oscillations [last term in (\ref{eq:largetimes1})] behaves
like $\lambda^2/\omega_0$ and the quantities
(\ref{eq:Zz})-(\ref{eq:Cchi}) become both unity.

In conclusion, only the exponential law survives in the limit
(\ref{eq:vHlimit}), with the correct normalization factor
($\cZ=1$), and one is able to derive a purely exponential behavior
(Markovian dynamics) from the quantum mechanical Schr\"{o}dinger
equation (unitary dynamics). It is important to notice that, in
order to obtain the exponential law, a normalizable state (such as
a wave packet) must be taken as initial state. Our initial state
$|2;0\rangle$ is indeed normalizable: see (\ref{eq:normaliz}).

\subsection{Two-level atom in the rotating-wave approximation}
\label{sec-VH2lev}
\andy{VH2lev}

Let us now proceed to a more formal analysis. Write the evolution
operator as
\andy{survFT}
\beq
\label{eq:survFT}
U(t)=\frac{i}{2\pi}\int_{\rm C}dE\frac{e^{-iEt}}{E-H},
\eeq
where the path C is a straight horizontal
line just above the real axis (this is the equivalent of the Bromwich path in
the Laplace plane). By defining the resolvents ($\Im E>0$)
\beq
S(E)\equiv\langle 2;0|\frac{1}{E-H_0}|2;0\rangle=\frac{1}{E-\omega_0},
\qquad
S'(E)\equiv\langle 2;0|\frac{1}{E-H}|2;0\rangle,
\eeq
Dyson's resummation reads
\beq
S'(E)=S(E)+\lambda^2S(E)\Sigma^{(2)}(E)S(E)+\lambda^4S(E)\Sigma^{(2)}(E)S(E)
\Sigma^{(2)}(E)S(E)+\dots
\eeq
where $\Sigma^{(2)}(E) = \langle 2;0|V(E-H_0)^{-1}V|2;0\rangle$ is
the 1-particle irreducible self-energy function. In the
rotating-wave approximation $\Sigma^{(2)}(E)$ consists only of a
second order diagram and can be evaluated exactly:
\beq
\Sigma^{(2)}(E)\equiv\int_0^\infty d\omega\frac{|\varphi(\omega)|^2}{E-\omega}
=i\Lambda Q(-i E/\Lambda),
\eeq
where the matrix element $\varphi=\varphi_{\bar{\beta}}$ in
(\ref{eq:phidef1}) and $Q$ is the function in (\ref{eq:qs4}). In
the complex $E$-plane $\Sigma^{(2)}(E)$ has a branch cut running
from 0 to $\infty$, a branching point in the origin and no
singularity on the first Riemann sheet. Summing the series
\andy{sumser}
\beq
S'(E)=\frac{1}{S(E)^{-1}-\lambda^2\Sigma^{(2)}(E)}=
\frac{1}{E-\omega_0-\lambda^2\Sigma^{(2)}(E)},
\label{eq:sumser}
\eeq
we obtain for the survival amplitude
\andy{survE}
\barr
\As(t)&\equiv&\langle2;0|e^{i H_0 t} U(t)|2;0\rangle=
\frac{i}{2\pi}\int_{\rm C}dE e^{-iEt} S'(E+\omega_0)\nonumber\\
     &=& \frac{i}{2\pi}\int_{\rm C}dE\frac{e^{-iEt}}
{E-\lambda^2\Sigma^{(2)}(E+\omega_0)} .
\label{eq:survE}
\earr
In Van Hove's limit one looks at the evolution of the system over
time intervals of order $t=\tilde t/\lambda^2$ ($\tilde t$
independent of $\lambda$), in the limit of small $\lambda$. Our
purpose is to see how this limit works in the complex-energy plane,
i.e.\ what is the limiting form of the propagator. To this end, by
rescaling time $\tilde t\equiv
\lambda^2 t$, we  can write
\andy{survEsc}
\beq
\As\left(\frac{\tilde{t}}{\lambda^2}\right)=
\frac{i}{2\pi}\int_{\rm C}
d\tilde E\frac{e^{-i\tilde{E}\tilde t}} {\tilde
E-\Sigma^{(2)}(\lambda^2\tilde E+\omega_0)},
\label{eq:survEsc}
\eeq
where we are naturally led to introduce the rescaled energy
$\widetilde E\equiv E/\lambda^2$. Taking the Van Hove limit we get
\andy{survEsclim}
\beq
\widetilde \As(\tilde{t})\equiv
\lim_{\lambda\to0}\As\left(\frac{\tilde{t}}{\lambda^2}\right)
=\frac{i}{2\pi}\int_{\rm C}
d\tilde E e^{-i\tilde{E}\tilde t}\widetilde S'(\tilde E).
\label{eq:survEsclim}
\eeq
\begin{figure}[t]
\epsfig{file=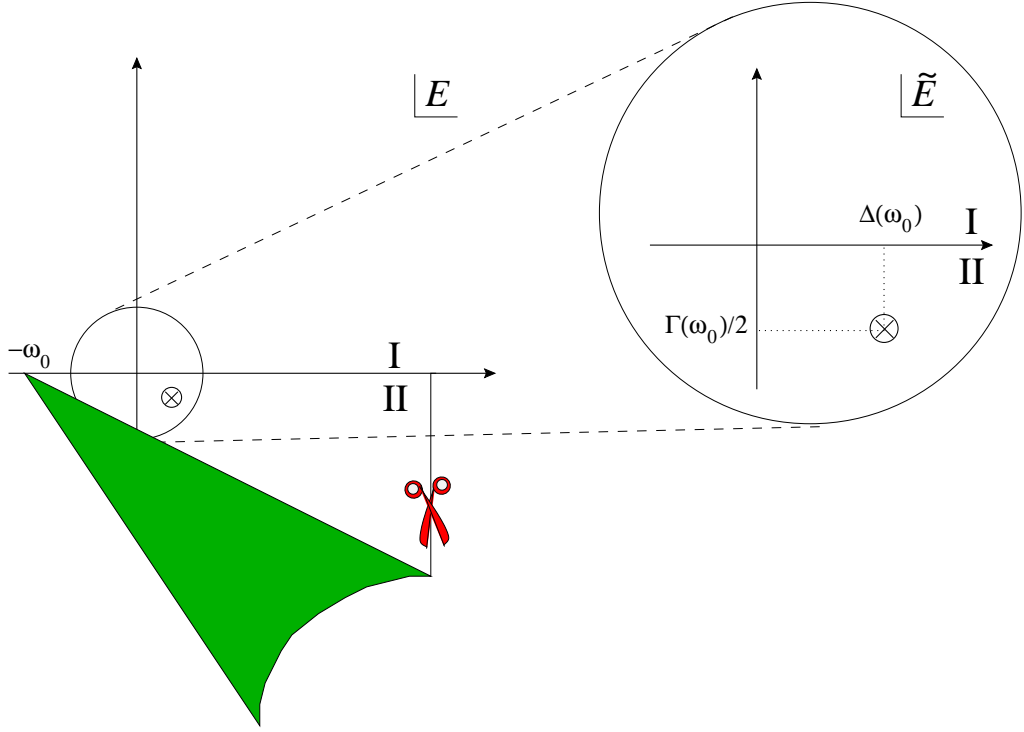, width=13.5cm}
\caption{Singularities of the propagator (\ref{eq:survE}) in the complex-$E$
plane. The first Riemann sheet ({\rm I}) is singularity free. The
logarithmic cut is due to $\Sigma^{(2)}(E)$ and the pole is located
on the second Riemann sheet ({\rm II}). In the complex-$\tilde E$
plane, the pole has coordinates
(\ref{eq:polecoord1})-(\ref{eq:polecoord2}). }
\label{fig:cmplxe}
\andy{sigma4}
\end{figure}
where the propagator in the rescaled energy reads
\andy{rescS}
\beq
\widetilde S'(\tilde E)\equiv\lim_{\lambda\to0}
\frac{1}{\tilde E-\Sigma^{(2)}(\lambda^2\tilde E+\omega_0)}
=\frac{1}{\tilde E-\Sigma^{(2)}(\omega_0+i0^+)},
\label{eq:rescS}
\eeq
the term $+i0^+$ being due to the fact that $\Im\tilde E>0$. The
self-energy function in the $\lambda \to 0$ limit becomes
\andy{sigmaomega0}
\beq
\label{eq:sigmaomega0}
\Sigma^{(2)}(\omega_0+i0^+)=-\int_0^\infty d\omega\frac{|\varphi(\omega)|^2}
{\omega-\omega_0-i0^+}=\Delta(\omega_0)-\frac{i}{2}\Gamma(\omega_0)
\eeq
where
\andy{polecoord1,2}
\barr
\Delta(\omega_0)&\equiv& {\cal P}\int_0^\infty d\omega\frac{|\varphi(\omega)|^2}
{\omega_0-\omega},
\label{eq:polecoord1} \\
\Gamma(\omega_0)&\equiv& 2\pi|\varphi(\omega_0)|^2,
\label{eq:polecoord2}
\earr
which yields a purely exponential decay (Weisskopf-Wigner
approximation and Fermi Golden Rule). In Figure \ref{fig:cmplxe} we
endeavoured to clarify the role played by the time-energy rescaling
in the complex-$E$ plane.

One can get a more detailed understanding of the mechanisms that
underpin the limiting procedure by looking at higher order terms in
the coupling constant. The pole of the original propagator
(\ref{eq:sumser}) satisfies the equation
\beq
E_{\rm pole}-\lambda^2\Sigma^{(2)}(E_{\rm pole}+\omega_0)=0 ,
\eeq
which can be solved by expanding the self-energy function
around $E=\omega_0$ in power series
\beq
\Sigma^{(2)}(E+\omega_0)=\Sigma^{(2)}(\omega_0)+ E \Sigma^{(2)'}(\omega_0)
+\frac{E^2}{2} \Sigma^{(2)''}(\omega_0) + \dots,
\eeq
whose radius of convergence is $\omega_0$, due to the
branching point of $\Sigma^{(2)}$ in the origin. We get (iteratively)
\beq
E_{\rm pole}=\lambda^2 \Sigma^{(2)}(\omega_0)+\lambda^4 \Sigma^{(2)'}
(\omega_0)\Sigma^{(2)}(\omega_0)+{\rm O}(\lambda^6),
\eeq
which, due to (\ref{eq:sigmaomega0}), becomes
\andy{poledef}
\beq
\label{eq:poledef}
E_{\rm pole}\equiv \Delta E-\frac{i}{2}\gamma=\lambda^2\Delta(\omega_0)-
i\frac{\lambda^2}{2}\Gamma(\omega_0)+ {\rm O}(\lambda^4).
\eeq
In the rescaled energy (\ref{eq:poledef}) reads
\beq
\tilde E_{\rm pole}=\frac{E_{\rm pole}}{\lambda^2}=\Delta(\omega_0)-
\frac{i}{2}\Gamma(\omega_0)+ {\rm O}(\lambda^2)\stackrel{\lambda\rightarrow 0}
{\longrightarrow}\Delta(\omega_0)-\frac{i}{2}\Gamma(\omega_0),
\eeq
which is the same as (\ref{eq:sigmaomega0}).
This is again the Fermi Golden Rule.

\subsection{$N$-level atom with counter-rotating terms }
\label{sec-VHNlev}
\andy{VHNlev}

\begin{figure}
\epsfig{file=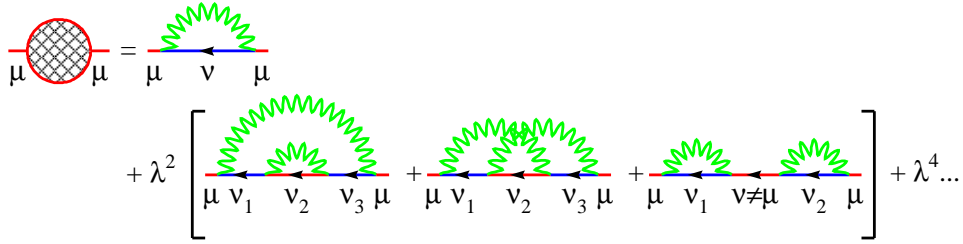, width=13.5cm}
\caption{Graphic representation of (\ref{eq:SEF}):
$\Sigma^{(2)}$ and $\Sigma^{(4)}$ are in the first and second line,
respectively.}
\label{fig:sigma4}
\andy{sigma4}
\end{figure}

Before proceeding to a general analysis it is interesting to see
how the above model is modified by the presence of the other atomic
levels and the inclusion of counter-rotating terms in the
interaction Hamiltonian. This will enable us to pin down other
salient features of the $\lambda^2t$ limit. The Hamiltonian is
\andy{newHV}
\beq
H=H_0^\prime+\lambda V^\prime,
\label{eq:newHV}
\eeq
where
\andy{newH0,V}
\barr
H_0^\prime & \equiv & \sum_\nu \omega_\nu b^\dagger_\nu b_\nu + \sum_\beta
\int_0^\infty d\omega \,
\omega a^\dagger_{\omega\beta} a_{\omega\beta} , \label{eq:newH0} \\
 V^\prime & = & \sum_{\mu,\nu} \sum_\beta \int_0^\infty
d\omega \left[ \varphi^{\mu\nu}_\beta(\omega) b^\dagger_\mu b_{\nu}
a^\dagger_{\omega \beta}
+ \varphi^{\mu\nu*}_\beta(\omega) b^\dagger_{\nu} b_\mu a_{\omega \beta}
\right],
\label{eq:newV}
\earr
where $\nu$ runs over all the atomic states and $b^\dagger_\nu,
b_\nu$ and $a^\dagger_{\omega\beta}, a_{\omega\beta}$ satisfy
anticommutation and commutation relations, respectively. [The
Hamiltonian (\ref{eq:tothaml1})-(\ref{eq:tothaml3}) is recovered if
we set $\omega_2=\omega_0, \; \omega_1=0$ and neglect the
counter-rotating terms.] Starting from the initial state
$|\mu;0\rangle$, Dyson's resummation yields
\andy{NBorn}
\beq\label{eq:NBorn}
S'(E)=\frac{1}{S(E)^{-1}-\lambda^2\Sigma(E)}=
\frac{1}{E-\omega_\mu-\lambda^2\Sigma(E)}
\eeq
and the 1-particle irreducible self-energy function takes the form
\andy{SEF}
\beq
\Sigma(E)=\Sigma^{(2)}(E)+\lambda^2\Sigma^{(4)}(E)+\dots,
\label{eq:SEF}
\eeq
with
\andy{gensigma2}
\beq\label{eq:gensigma2}
\Sigma^{(2)}(E)\equiv\sum_{\nu,\beta}\int_0^\infty d\omega
\frac{|\varphi^{\nu\mu}_\beta(\omega)|^2}{E-\omega_\nu-\omega} .
\eeq
Both $\Sigma^{(2)}$ and $\Sigma^{(4)}$ are shown as Feynman diagrams in Figure
\ref{fig:sigma4}.
In the Van Hove limit one obtains
\andy{sigma2lim}
\beq
\Sigma(\lambda^2\tilde E+\omega_\mu)\stackrel{\lambda\rightarrow 0}
{\longrightarrow}\Sigma^{(2)}(\lambda^2\tilde E+\omega_\mu)\Big|_
{\lambda=0}=\Sigma^{(2)}(\omega_\mu+i0^+).
\label{eq:sigma2lim}
\eeq
The propagator in the rescaled energy takes now the form
\beq
\widetilde S'(\tilde E)=\lim_{\lambda\to0}
\frac{1}{\tilde E-\Sigma^{(2)}
(\lambda^2\tilde E+\omega_\mu) +{\rm O}(\lambda^2)}
=\frac{1}{\tilde E-\Sigma^{(2)}(\omega_\mu+i0^+)},
\eeq
where
\beq
\Sigma^{(2)}(\omega_\mu+i0^+)=\sum_{\nu,\beta}\int_0^\infty d\omega
\frac{|\varphi^{\nu\mu}_\beta(\omega)|^2}{\omega_\mu-\omega_\nu-\omega+i0^+}.
\eeq
The last two equations correspond to (\ref{eq:rescS})-(\ref{eq:sigmaomega0}):
the propagator reduces to that of a generalized rotating-wave approximation.

We see that the Van Hove limit works by following two logical
steps. First, it constrains the evolution in a Tamm-Dancoff sector:
the system can only ``explore" those states that are directly
related to the initial state $\mu$ by the interaction $V'$. In
other words, in this limit, the ``excitation number" ${\cal N}_\mu
\equiv b^\dagger_\mu b_\mu + \sum_{\beta, \omega}
a^\dagger_{\omega\beta} a_{\omega\beta}$ becomes a conserved
quantity (even though the original Hamiltonian contains
counter-rotating terms) and, as a consequence, the self-energy
function consists only of a second order contribution that can be
evaluated exactly. Second, it reduces this second order
contribution, which depends on energy as in (\ref{eq:gensigma2}),
to a constant (its value in the energy $\omega_\mu$ of the initial
state), like in (\ref{eq:sigma2lim}). Hence the analytical
properties of the propagator, which had branch-cut singularities,
reduce to those of a single complex pole, whose imaginary part
(responsible for exponential decay) yields the Fermi Golden rule,
evaluated at second order of perturbation theory.

Notice that it is the latter step (and not the former one) which is
strictly necessary to obtain a dissipative behavior: Indeed,
substitution of the pole value in the total self-energy function
yields exponential decay, including, as is well known, higher-order
corrections to the Fermi Golden Rule. On the other hand, the first
step is very important when one is interested in computing the
leading order corrections to the exponential behavior. To this
purpose one can solve the problem in a restricted Tamm-Duncoff
sector of the total Hilbert space (i.e., in an eigenspace of ${\cal
N}_\mu$ ---
in our case, ${\cal N}_\mu=1$) and exactly evaluate the
evolution of the system with its deviations from exponential law.

Let us add a final remark. As is well known, a nondispersive
propagator yields a Markovian evolution. Let us briefly sketch how
this occurs in the present model. From (\ref{eq:NBorn}),
antitransforming,
\beq
\frac{i}{2\pi}\int_C dE e^{-iEt} \left(E S'(E+\omega_\mu)-1\right)=
\frac{i}{2\pi}\int_C dE e^{-iEt}
\lambda^2\Sigma (E+\omega_\mu) S'(E+\omega_\mu),
\eeq
we obtain (for $t>0$)
\andy{nonlocal}
\beq
i\dot \As(t)=\lambda^2\int_0^t d\tau
\sigma(t-\tau)\As(\tau) ,
\label{eq:nonlocal}
\eeq
where $\As(t)$ is the survival amplitude (\ref{eq:survampl}) and
\beq
\sigma(t)\equiv\frac{1}{2\pi}\int_C dE e^{-iEt}\Sigma(E+\omega_\mu)
=\frac{e^{i\omega_\mu t}}{2\pi}\int_C dE e^{-iEt}\Sigma(E).
\eeq
Equation (\ref{eq:nonlocal}) is clearly nonlocal in time and all
memory effects are contained in  $\sigma(t)$, which is the
antitransform of the self-energy function. If such a self-energy
function is a complex constant (energy independent), $\Sigma(E)=C$,
then $\sigma(t)=C\delta(t)$ and equation (\ref{eq:nonlocal})
becomes
\beq
i\dot \As(t)=\lambda^2 C \As(t),
\eeq
describing a Markovian behavior, without memory effects
\cite{VanKampen}. In particular, the Van Hove limit is equivalent
to set $C=\Sigma^{(2)}(\omega_\mu+i0^+)$ and the Weisskopf-Wigner
approximation is $C=\Sigma^{(2)}(\omega_\mu+i0^+)+{\rm
O}(\lambda^2)$.

In conclusion, in the Van Hove limit, the evolution of our system,
which was nonlocal in time due to the dispersive character of the
propagator (the self-energy function depended on $E$) becomes local
and Markovian (only the value of the self-energy function in
$\omega_\mu$ determines the evolution).

\setcounter{equation}{0}
\section{General framework}
\label{sec-general}
\andy{general}
We can now further generalize our analysis: consider the
Hamiltonian
\beq
H=H_0+\lambda V
\eeq
and suppose that the initial state $|a\rangle$ has the following
properties
\andy{aprop}
\barr
& & H_0|a\rangle=E_a|a\rangle,\qquad
\langle a|V|a\rangle=0, \nonumber\\
& & \langle a|a\rangle=1.
\label{eq:aprop}
\earr
The survival amplitude of state $|a\rangle$ reads
\andy{survEgen}
\barr
\As(t)&\equiv&\langle a|e^{i H_0 t} U(t)|a\rangle=
\frac{i}{2\pi}\int_{\rm C}dE e^{-iEt} S'(E+E_a)\nonumber\\
&=&
\frac{i}{2\pi}\int_{\rm C}dE\frac{e^{-iEt}}
{E-\lambda^2\Sigma(E+E_a)},
\label{eq:survEgen}
\earr
where $S'(E)\equiv\langle a|(E-H)^{-1}|a\rangle$ and $\Sigma(E)$ is
the 1-particle irreducible self-energy function, that can be
expressed by a perturbation expansion
\andy{sigmagen}
\beq
\lambda^2 \Sigma(E) = \lambda^2 \Sigma^{(2)}(E) + \lambda^4 \Sigma^{(4)}(E)
+\cdots .
\label{eq:sigmagen}
\eeq
The second order contribution has the general form
\andy{sigma2gen}
\barr
\Sigma^{(2)}(E)&\equiv&\langle a|V P_d \frac{1}{E-H_0} P_d V|a\rangle
=\sum_{n\neq a} \left| \langle a|V|n\rangle \right|^2\frac{1}{E-E_n}
\nonumber\\
&=& \int_0^\infty \frac{dE'}{2\pi} \frac{\Gamma(E')}{E-E'},
\label{eq:sigma2gen}
\earr
where $P_d=1-|a\rangle\langle a|$ is the projector over the decayed
states, $\{|n\rangle\}$ is a complete set of eingenstates of $H_0$
($H_0|n\rangle=E_n|n\rangle$ and we set $E_0=0$) and
\andy{genGamma}
\beq
\Gamma(E)\equiv2\pi\sum_{n\neq a}\left| \langle a|V|n\rangle \right|^2\delta(E-E_n).
\label{eq:genGamma}
\eeq
Notice that $\Gamma(E)\geq0$ for $E>0$ and is zero otherwise. In
the Van Hove limit we  get
\andy{gensurvEsclim}
\beq
\widetilde \As(\tilde{t})\equiv
\lim_{\lambda\to0}\As\left(\frac{\tilde{t}}{\lambda^2}\right)
=\frac{i}{2\pi}\int_{\rm C}
d\tilde E e^{-i\tilde{E}\tilde t}\widetilde S'(\tilde E),
\label{eq:gensurvEsclim}
\eeq
where the resulting propagator in the rescaled energy
$\tilde E=E/\lambda^2$ reads
\andy{genrescS}
\beq
\widetilde S'(\tilde E)
=\frac{1}{\tilde E-\Sigma^{(2)}(E_a+i0^+)}.
\label{eq:genrescS}
\eeq
To obtain this result we used
\andy{gensigma2lim}
\beq
\Sigma(\lambda^2\tilde E+E_a)\stackrel{\lambda\rightarrow 0}
{\longrightarrow}\Sigma^{(2)}(\lambda^2\tilde E+E_a)\Big|_
{\lambda=0}=\Sigma^{(2)}(E_a+i0^+)
\label{eq:gensigma2lim}
\eeq
(Weisskopf-Wigner approximation and Fermi Golden Rule).

\begin{figure}[t]
\epsfig{file=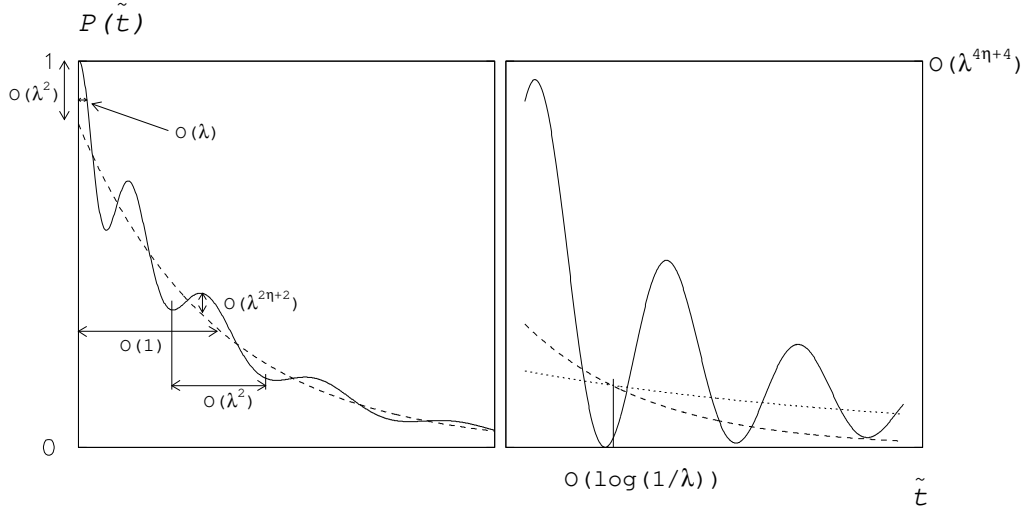, width=13.5cm}
\caption{Essential features (not in scale!) of the survival probability
as a function of the rescaled time $\tilde t$.
The Zeno time is ${\rm O}(\lambda)$,
the lifetime ${\rm O}(1)$, during the whole evolution there are
oscillations of amplitude ${\rm O}(\lambda^{2\eta+2})$ and the transition
to a power law occurs after a time ${\rm O}(\log (1/\lambda))$ [see
(\ref{eq:resctauZ})-(\ref{eq:resctaupow})]. From (\ref{eq:genZz}),
the normalization factor becomes unity like $1-{\rm O}(\lambda^2)$.
The dashed line is the exponential and the dotted line the power
law.}
\label{fig:summa}
\andy{summa}
\end{figure}

Just above the positive real axis we can write
\andy{sigmaEa}
\beq
\label{eq:sigmaEa}
\Sigma^{(2)}(E+i0^+)=\Delta(E)-\frac{i}{2}\Gamma(E),
\eeq
where
\andy{genpolecoord1}
\beq
\Delta(E) = {\cal P}\int_0^\infty \frac{dE'}{2\pi}\frac{\Gamma(E')}{E-E'}.
\label{eq:genpolecoord1} \\
\eeq
Let $\Gamma(E)$ be sommable in $(0,+\infty)$. Then
\andy{Gammalim}
\beq
\Gamma(E)\propto E^{\eta-1}\quad\mbox{for}\quad E\to0,
\label{eq:Gammalim}
\eeq
for some $\eta>0$, and one gets the following asymptotic behavior
at short and long times:
\andy{genshortt,genlargetimes1}
 \barr
\!\! P(t) \! &\sim& \!
1 - \frac{t^2}{\tau_{\rm Z}^2} \qquad (t\ll \tau_{\rm Z}),
 \label{eq:genshortt} \\
\!\! P(t) \! &\sim& \!
|Z|^2 e^{-t/\tau_{\rm E}}+ \lambda^4\frac{|C|^2}{(E_a t)^{2\eta}}
+ 2 \lambda^2\frac{|CZ|}{(E_a t)^\eta} e^{-t/2\tau_{\rm E}}
 \cos\left[(E_a+\Delta E) t -\zeta \right] \label{eq:genlargetimes1} \nonumber\\
  & & \qquad\qquad\qquad\qquad\qquad\qquad\qquad\qquad(t\gg \tau_{\rm Z}),
 \earr
where
 \andy{gentauZ,genfgr,genshift,genzeta,genZz,genCchi}
 \barr
 \tau_{\rm Z} &=& \frac{1}{\lambda} \left[\int_0^\infty
 \frac{dE}{2\pi}\Gamma(E)\right]^{-1/2},
 \label{eq:gentauZ}\\
 \tau_{\rm E} &=& \frac{1}{\lambda^2 \Gamma(E_a)},
 \label{eq:genfgr}\\
\Delta E &=&\lambda^2 \Delta(E_a),
 \label{eq:genshift} \\
\zeta&=& {\rm Arg} Z - {\rm Arg} C,
\label{eq:genzeta}\\
Z &=& 1 + {\rm O}(\lambda^2),
 \label{eq:genZz} \\
C &=& 1 + {\rm O}(\lambda^2).
 \label{eq:genCchi}
\earr
The transition to a power law occurs when the first two terms in
the r.h.s. of (\ref{eq:genlargetimes1}) are comparable, namely for
$t=\tau_{\rm pow}$, where $\tau_{\rm pow}$ is solution of the
equation
\andy{tpoweq}
\beq
\frac{\tau_{\rm pow}}{\tau_{\rm E}}=4(\eta+1)\log\frac{1}{\lambda}
+2\eta\log\frac{E_a}{\Gamma(E_a)}+\log\left|\frac{Z}{C}\right|
+\eta\log\frac{\tau_{\rm pow}}{\tau_{\rm E}},
\label{eq:tpoweq}
\eeq
i.e., for $\lambda\to 0$
\andy{gentaupow}
\beq
\tau_{\rm pow} = 4 \tau_{\rm E} (\eta+1) \log\lambda^{-1}
+{\rm O}\left(\log\log \lambda^{-1}\right).
\label{eq:gentaupow}
\eeq
Let us now look at the temporal behavior for a small but {\em
finite} value of $\lambda$, using Van Hove's technique. In the
rescaled time, $\tilde t=\lambda^2 t$, the Zeno region vanishes
\andy{resctauZ}
\beq
\tilde\tau_{\rm Z}\equiv\lambda^2\tau_{\rm Z}=\lambda\left[\int_0^\infty
 \frac{dE}{2\pi}\Gamma(E)\right]^{-1/2}={\rm O}(\lambda)
\label{eq:resctauZ}
\eeq
and Eq.\ (\ref{eq:genlargetimes1}) becomes valid at {\em shorter
and shorter} (rescaled) times and reads
\andy{genlargetimes2}
\barr
& & P(\tilde t) \sim |Z|^2 e^{-\tilde t/\tilde\tau_{\rm E}} +
\lambda^{4(\eta+1)}\frac{|C|^2}{(E_a \tilde t)^{2\eta}} \nonumber\\
& &\quad\quad\quad\quad\quad + 2
\lambda^{2(\eta+1)}\frac{|CZ|}{(E_a t)^\eta} e^{-\tilde
t/2\tilde\tau_{\rm E}}
 \cos\left(\frac{E_a+\Delta E}{\lambda^2} \tilde t -\zeta \right),
\label{eq:genlargetimes2}
\earr
where
\andy{resctauE, resctaupow}
\barr
\tilde\tau_{\rm E} &\equiv& \lambda^2\tau_{\rm E}=
\frac{1}{\Gamma(E_a)}={\rm O}(1),
\label{eq:resctauE}\\
\tilde\tau_{\rm pow}&\equiv& \lambda^2\tau_{\rm pow}\simeq
4 \tilde\tau_{\rm E} (\eta+1) \log\frac{1}{\lambda}
={\rm O}\left( \log\frac{1}{\lambda} \right).
\label{eq:resctaupow}
\earr
Figure \ref{fig:summa} displays the main features of the temporal
behavior of the survival probability.
The typical values of the physical constants [see for instance
(\ref{eq:lambdachi})] yield very small deviations from the
exponential law. For this reason, we displayed in Figure
\ref{fig:summa} the survival probability  by greatly exaggerating
its most salient features.

The Van Hove limit performs several actions at once: It makes the
initial quadratic (quantum Zeno) region vanish, it ``squeezes" out
the oscillations and it ``pushes" the power law to infinity,
leaving only a clean exponential law at all times, with the right
normalization. All this is not surprising, being implied by the
Weisskopf-Wigner approximation. However, the concomitance of these
features is so remarkable that one cannot but wonder at the
effectiveness of this limiting procedure. In atomic and molecular
physical systems the smallness of the coupling constant and other
physical parameters makes the experimental observation of
deviations from exponential a very difficult task (see for example
the simple model investigated in Section 2). The eventuality that
alternative physical systems might exhibit experimentally
observable non-exponential decays, as well as the possibility of
modifying the lifetimes of unstable systems by means of intense
laser beams \cite{MPS} are at present under investigation.


%
%



\end{document}